\documentclass[a4paper,twoside]{article}

\usepackage{epsfig}
\usepackage{subcaption}
\usepackage{calc}
\usepackage{amssymb}
\usepackage{amstext}
\usepackage{amsmath}
\usepackage{amsthm}
\usepackage{multicol}
\usepackage{pslatex}
\usepackage{apalike}
\let\bibhang\relax
\usepackage{natbib} 

\usepackage{array}
\usepackage{tabularx}
\usepackage[utf8]{inputenc}  
\usepackage[T1]{fontenc}  
\usepackage{graphicx}
\usepackage{wrapfig}
\usepackage[labelfont=bf]{caption}

\usepackage[table]{xcolor}
\usepackage{multicol}
\usepackage{hyperref}

\usepackage{SCITEPRESS}     

\newcommand{\wrt}{\textit{wrt}. }

\newcommand{\ie}{\textit{i}.\textit{e}., }

\newcommand{\remove}[1]{}

 \definecolor{lightgray}{gray}{0.95}

\begin{document}

\title{A Malaria Control Model Using Mobility Data : An Early Explanation of Kedougou's Case in Senegal}

\author{\authorname{Lynda BOUZID KHIRI\sup{1}, Ibrahima GUEYE\sup{3}, Hubert NAACKE\sup{1}, Idrissa SARR\sup{2} and Stéphane GANCARSKI\sup{1}}
\affiliation{\sup{1}Sorbonne Université, CNRS, Laboratoire d’Informatique de Paris 6, LIP6, F-75005}
\affiliation{\sup{2}Cheikh Anta Diop University, Department of Mathematics and Computer Science,	Dakar - Fann BP, 5005, Senegal}
\affiliation{\sup{2}Ecole Polytechnique de Thiès, LTISI; Sénégql}
\email{\{hubert.naacke, stephane.gancarski\}@lip6.fr, idrissa.sarr@ucad.edu.sn, igueye@ept.sn}
}


\abstract{Studies in malaria control cover many areas such as medicine, sociology, biology, mathematic, physic, computer science and so forth. Researches in the realm of mathematic are conducted to predict the occurrence of the disease and to support the eradication process. Basically, the modeling methodology is predominantly deterministic and differential equation based while selecting clinical and biological features that seem to be important. Yet, if the individual characteristics matter when modeling the disease, the overall estimation of the malaria is not done based on the health status of each individual but in a non-specified percentage of the global population. 
The goal of this paper is to propose a model that relies on a daily evolution of the individual's state, which depends on their mobility and the characteristics of the area they visit.  Thus, the mobility data  of a single person moving from one area to another, gathered thanks to mobile networks,  is the essential building block  to predict the outcome of the disease. We implement our solution and demonstrate its effectiveness through empirical experiments. The results show how promising the model is in providing possible insights into the failure of the disease control in the Kedougou region.}

\onecolumn \maketitle \normalsize \setcounter{footnote}{0} \vfill

\section{Introduction}
\label{sec:intro}

	Human malaria is caused by infection by the \textit{Plasmodium falciparum} and four other species of parasites, leading to almost 600,000 deaths and 100--250M febrile episodes annually~\cite{Who2016}. 
Even though the disease has been investigated for hundreds of years, it still remains a major public health problem in Sub-Saharan Africa (SSA)  where 90\% of malaria cases and deaths are approximately reported in 2017~\cite{Who2016}.

Many SSA countries have set the goal of eliminating malaria for the upcoming decades outbreaks~\cite{Ruktanonchai2016}. Among these countries, Senegal has initiated its  National Program Against Malaria (PNLP)~\cite{sn:gouv:sante}. 
Besides a weekly follow-up of the disease evolution, the PNLP has allowed to intensify the
coverage of key malaria interventions over the country in terms of impregnated mosquito nets, insecticide (ITN), indoor residual spraying, preventive treatment by
intermittent administration to women intestines (TPI), rapid diagnostic tests (RDTs) and therapeutic combinations based on Artemisinin (CTA) \cite{thiam2011}. Those strategies have lowered the malaria incidence (relative number of infected people for 1000 inhabitants) to a relative small number estimated to 25 in 2017. However, in the southeastern part of the country (Kedougou region  alongside Kolda and Tambacounda) accounts for 75\% of malaria cases and 45\% of malaria-related deaths. Specially,
the malaria incidence was estimated to 429 in 2017 for Kedougou while the other regions of Senegal had an average incidence below 10. Such a situation explains the negative impact on Kedougou of the overall strategies taken to face the disease and why it states as a serious problem for malaria pre-elimination in the country. In this work, we rely on Kedougou to show forgotten aspects in antimalarial policies and that one should consider for more efficient actions.
\begin{figure}
\centering
\includegraphics[width=2in]{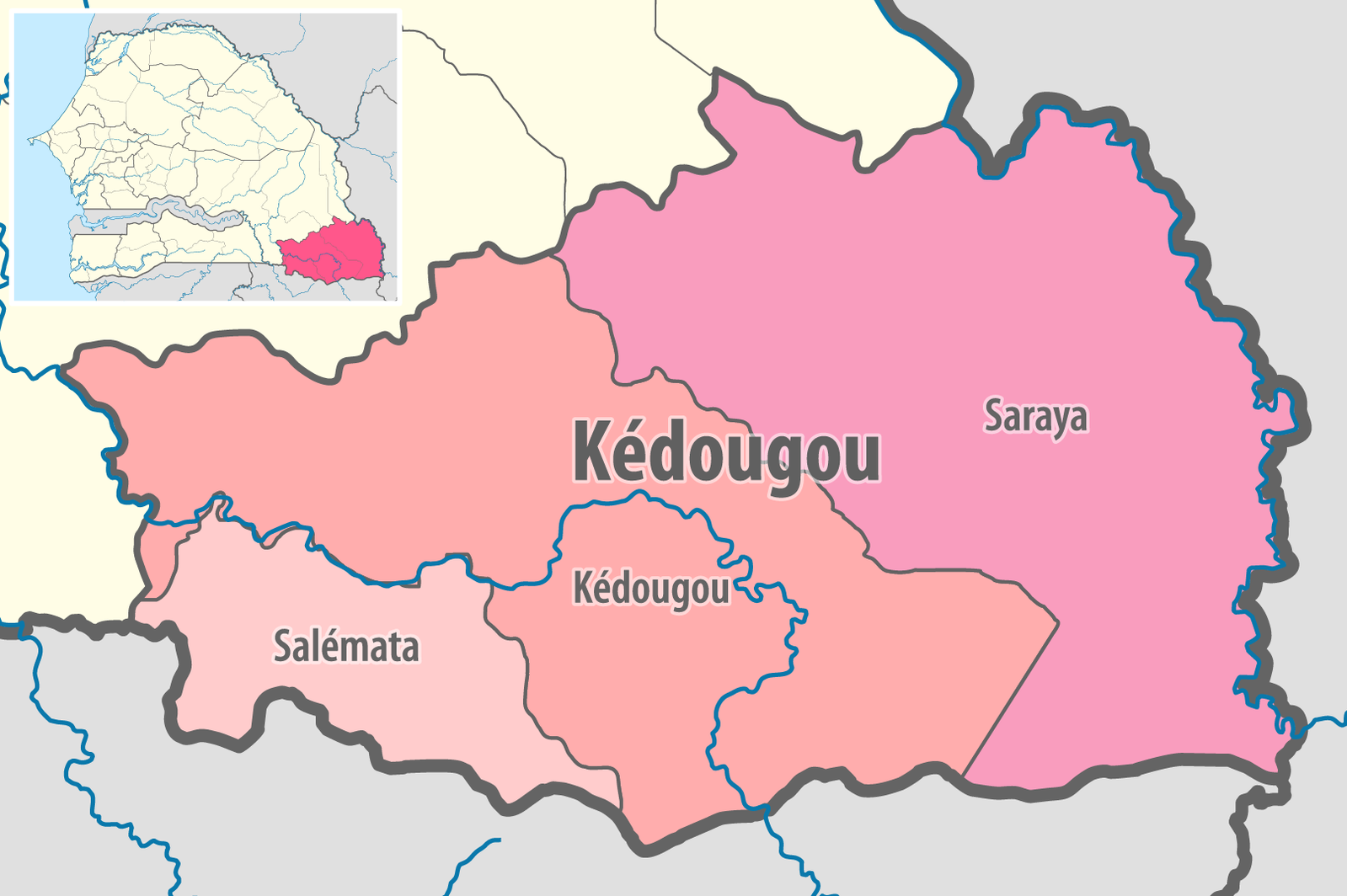}
\caption{Kedougou region with its 3 departments}
\label{kedo}
\end{figure}

Actually, Kedougou is the largest city in southeastern Senegal and located near to 700 km from the capital Dakar. 
It has a dry tropical climate, with average annual rainfall over 1000 mm, spread over a rainy season that lasts from May to November. It is a mainly agricultural region with the cultivation of cereals (rice, corn, sorghum, millet, fonio ...) and many forest fruit species including mango, shea, palm, etc. Moreover, it offers a variety of natural attractions including those of the Niokolo Koba national park, the hills where trekking activities are practiced and the Dindefelo waterfall. The discovery of deposits of uranium, granites, marble and other ornamental rocks, but also industrial minerals such as phosphate and kaolin ranks Kedougou as a cornerstone  mining region. All these characteristics, along with its proximity to Mali and Guinea make Kedougou a true crossroads over all the year, which leads to a sustained human mobility rate. As shown in the Figure \ref{kedo}, the region of Kedougou is divided into three departments, namely, Salemata with 14.6 \% of the population, Kedougou department  that shelters 51.9\% and Saraya, 33.5\% \cite{sn:ansd:2013}. As depicted on the map, Kedougou department is on the center of the region and hosts the main infrastructures such as markets, health centers, and so on. This geo-administrative division raises an intra-mobility rate of individuals within Kedougou region.

As a conclusion, Kedougou region is characterized by two types of mobility : an intra mobility for daily or weekly needs of permanent residents, and an extra mobility at both of the country and the west African community level. 
Our goal is to provide tools highlighting the negative impact of this mobility on the malaria incidence in the region. 

\begin{figure}
\centering
\includegraphics[width=2in]{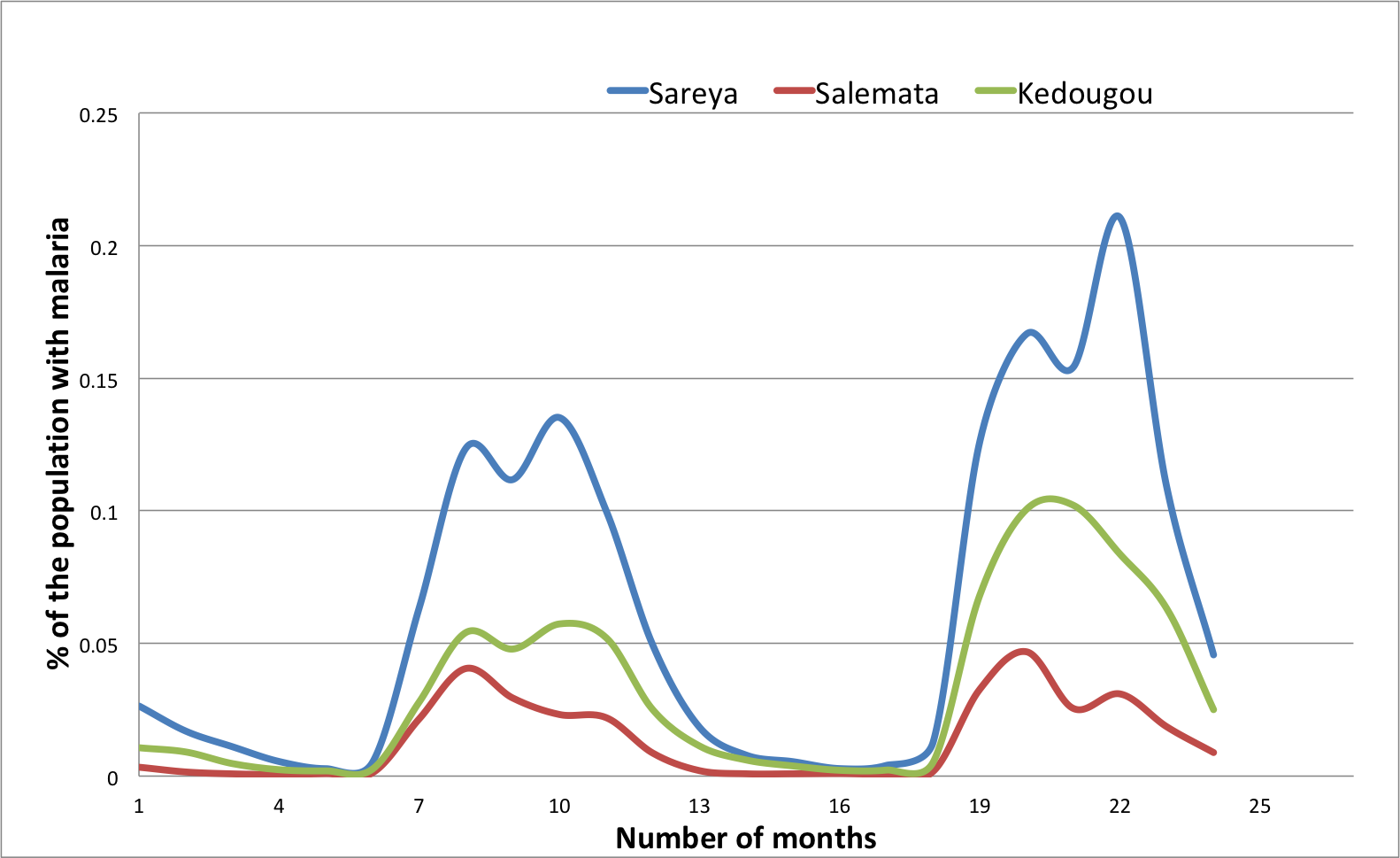}
\caption{The Variation of malaria between Jan. 2016 and Dec. 2017}
\label{kedmalaria}
\end{figure}

Some statistical data from the PNLP and related to Kedougou region \cite{sn:gouv:sante} are used to plot the Figure \ref{kedmalaria}, that shows the variation of new malaria cases over 24 months, \textit{i,e,.} from January 2016 to December 2017. 
The first observation is that the number of cases raises drastically just after the beginning of the rainfall season (sixth month of each year) and decreases with the end of the rainy annual period (ninth month of the year). 
This situation is  explained by the fact that mosquitoes population is multiplied during rainy season, and therefore, with a similar climatic conditions during two successive years, we see almost the same trends.    

However, when focusing on the different curves, we find out that the epidemic of the three departments are not similar. First of all, the different peaks of the malaria cases in each departments do not occur at the same time and are actually staggered by a few weeks. Moreover, we note that the epidemic in Saraya lasts longer than the ones in Kedougou or Salemata even though the three departments have similar climactic conditions. Thus,  rainfalls do not totally explain the spread of the epidemic over 6 months. Still, we know that there is a lot of movement towards the zone (trade and mining with other border countries) and it is demonstrated that human mobility has an  impact on malaria control and elimination \cite{Gharbi2013, Ruktanonchai2016} and even in malaria-free countries \cite{Srilanka2017}. Bearing this in mind, it makes sense to relate the epidemic outbreak of a given area to the arrival of outsiders who have been exposed in other areas during different periods.
Surprisingly, despite the sustained mobility around the region of Kedougou, the PNLP does not yet includes human movements in its control strategies.  

We make the assumption that the arrival of infected people in a given area makes  the  epidemic to last longer. The impact on the epidemic extension depends on the arrival date of new people and the epidemic  state of the area they come from. 
Our goal is to demonstrate that this assumption is plausible through a mathematical model. 
In fact, mathematical models have been frequently used as tools in malaria control \cite{Chitnis2006,Dimitrov2010,Mandal2011,Chitnis2008,Schoenstadt2017,Ruktanonchai2016,Greenwood1991,Gu2003,Koella1991,Filipe2007,RDC2019}. Existing models consider different parameters and aspects that influence the disease dispersal, such as heterogeneity, immunity, recovery time and more recently, human mobility. However, the models with human mobility only deal with the movement of people through a coarse grain approach, which assumes a global migration ratio from an area to another one. Rather, a finer grain approach can be used thanks to personalized geo-position information   (GPS coordinates) from mobile phones. Such a finer approach allows a better understanding of the disease evolution at each time, on each area. Therefore, it helps to determine the antimalarial actions in a more dynamic and efficient fashion. 

The main contributions of this paper can be summarized as follow:

\begin{itemize}
\item We define a mathematical model that takes into account individual mobility and immunity. To this end, we assume that we have real-time data from mobile devices allowing to establish the pattern mobility of everyone, and his(her) state wrt. the malaria (ill or not). Hence, we build a discrete model that mainly differs from existing works by the fact that the global status of a given space is obtained by aggregating the health status of each individual.

\item We implement a simulation software with respect to climatic conditions and human movements over the time. The software is designed so that the relevant parameters of the disease can be adjusted according to a real life situation of a given area. 

\item We conduct a set of experiments to validate our approach 
while we point out 
many benefits of our solution in terms of explaining the disease evolution in  areas like Kedougou.
To this end, we rely on synthetic data according to realistic scenarios since real-time data are not yet available. We show the impact of different factors (characteristics of areas, mobility and state of individuals) on the malaria propagation. This allows for measuring the impact of malaria control actions (eradication, prevention) in an accurate way, which helps deciding which actions should be prioritized.
\end{itemize}

\section{Background}
\label{back}

 Mathematical models have been used  to predict the occurrence of a disease and to control its dispersal. 
Basically, the modeling methodology is mainly deterministic and based on differential equations while selecting clinical and biological features that seem to be important~\cite{Greenwood1991,Chitnis2006,Chitnis2008,Schoenstadt2017,Ruktanonchai2016}.
The first models that were developed examine the interaction of human, vectors and parasites with a coarse granularity, for instance, at the city/country level~\cite{Mandal2011}. More recent models have attempted to handle heterogeneity such as the individual immunity  \cite{Gu2003,Filipe2007}, the space and contact network \cite{Parham2006}, the recovery rate \cite{GU200371}, etc. A recent work integrates human  mobility data \cite{Ruktanonchai2016} for explaining and eliminating the disease in a particular area. 

One of the first model, known as the classical "Ross model", was developed by Sir Ronald Ross, which explained the relationship between the number of mosquitoes and incidence of malaria in humans~\cite{Ross1911}. In such a model, the population is  divided into several {\em compartments}, which represent their health status regarding  the pathogen.  These status or compartments are represented by the standard notation  \textit{S-E-I-R}, based on the work presented in \cite{Kermack1927}. The \textit{S} class stands for the fraction of host population that is susceptible to infection, while the \textit{E}  category indicates the fraction of population whose individuals have been infected but are not yet infectious themselves due to a latency period. The \textit{I} class  represents infectious individuals who may  infect other individuals through interactions with mosquitoes. Finally, the \textit{R} class  portrays individuals who recover from the infection. Notice that sometimes, \textit{R} may include individuals who recover with temporary or permanent immunity. With these different classes insight, one  may observe eight possible models: \textit{SIS}, \textit{SEI}, \textit{SEIS}, \textit{SIR}, \textit{SIRS}, \textit{SEIR} and \textit{SEIRS}.
Note that both mosquito and host population may be related with these compartments in a malaria disease case. That is, the malaria transmission model is described along two aspects, one representing humans and the other representing mosquitoes. However, a mosquito can not recover from its infectiousness,
so its infective period ends with its death.

\section{Discrete Malaria Model}
\label{sec:model}

As we pointed out earlier, we aim at integrating user mobility information into a malaria transmission model. The reason behind this is that knowing the mobility  and state of each individuals allows for assessing the specific persons that diffuse the disease instead of finding a proportion of population as done by existing models.
 In this sense, our approach differs to others by the fact we estimate the probability of each individual to be part of one classes (SEIR), and therefore, we deduce the global population that belong to each class at each time unit.

\subsection{Global model overview}
\begin{figure}[htbp]
\centering
\includegraphics[width=\linewidth]{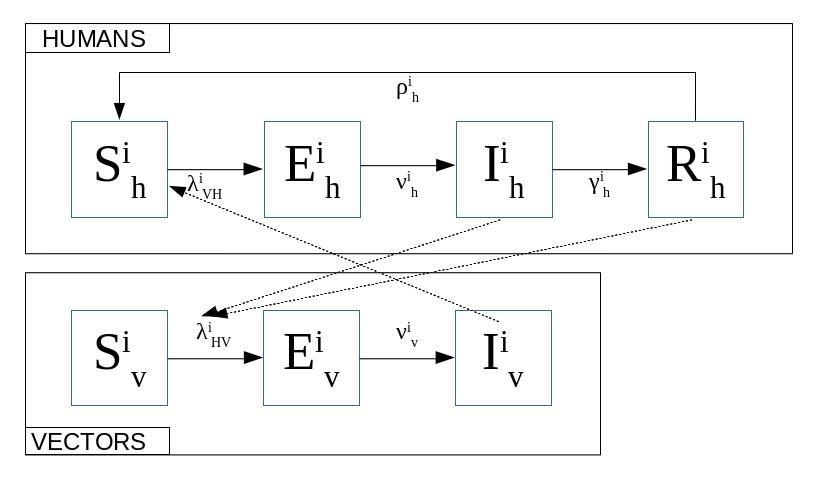}
\caption{Malaria model in patch $i$ \cite{Chitnis2008}.}
\label{sierstand}
\end{figure}

We assume a multi-patches area where each patch has a specific configuration to impact the malaria disease propagation. Individuals can move from one patch to another while mosquitoes are set to stay in only one patch.  
To model the transmission, we extend the \textit{SEIRS-SEI} model proposed in \cite{Chitnis2008} by introducing patches and individuals data mobility. Fig. \ref{sierstand}
shows the proposed 
malaria transmission model. Solid arrows denote intra-species progression into classes while dotted
arrows denote inter-species transmission.
With this model, for each patch $i$ the human population is divided into four classes: susceptible, $S_h^i$; exposed, $E_h^i$; infectious, $I_h^i$; and recovered (partially and/or temporary immune), $R_h^i$. Moreover, mosquitoes population is divided into three classes : susceptible $S_v^i$,  exposed $E_v^i$ , infectious $I_v^i$.  
We assume a constant population (\textit{i.e.}, birth rate equals the death rate). Initially, all individuals are in the susceptible class except a low percentage that live with the parasites. This situation is realistic in a context where the malaria parasite is still present. 
%
%
Basically, a proportion of the susceptible individuals that move from the $S$ to the $E$ class due to mosquito bites is characterized by the \emph{force of infection} (FoI) $\lambda^i_{vh}$.
Among exposed individuals, there is a proportion $\nu^i_h $ that enter to the infectious class. $\nu^i_h $ depends on a time period, called an intrinsic incubation period, which is linked to the parasite species (\textit{i.e.,  Plasmodium falciparum}). Later on (approximately a couple of weeks), part of  infectious human ($\gamma_h^i$) recover and join the $R$ class where there may acquire a certain immunity to the disease and do not get clinically ill. However, they still host low amount of parasites and can pass the infection to mosquitoes with a low rate. Over the time, the immunity of  individuals vanishes and leads a proportion of them  ($\rho_h^i$) to return to the $S$ class. 
Regarding  the mosquitoes population, the same flowchart is observed with only three classes. It is worth noting that the FoI of mosquitoes ($\lambda^i_{hv}$) differs to the FoI of human, so as the incubation period between mosquitoes and humans. 
The main parameters used to devise the model are structured into two  categories: patch parameters and individual parameters. In the following, we present a short overview of these parameters.

\subsection{Dealing with patch and individual characteristics}
\label{characteristics}
\subsubsection{Patch characteristics}
\begin{table*}[t]
\centering
\rowcolors{2}{lightgray}{white}
\renewcommand{\arraystretch}{0.8}
\begin{tabular}{|c|p{9.5cm}|}
\hline
\textbf{Parameter} & \textbf{Description} \\
\hline
$N_h^i(t)$ & Human population in patch $i$ at time $t$\\
\hline
$N_v^i(t)$ & Vector Population in patch $i$ at time $t$\\
\hline
$I_v^i(t)$ & Infected mosquitoes in patch $i$ at time $t$\\
\hline
$I_h^i(t)$ & Infected humans in patch $i$ at time $t$\\
\hline
$R_h^i(t)$ & Infected individuals in patch $i$ that recover at time $t$\\
\hline
$\beta^i_{hv}$  & Probability that an infectious person infects a susceptible mosquito during a contact within the patch $i$\\
\hline
$\beta^i_{vh}$ & Probability that an infectious mosquito infects a susceptible individual during a bite in patch $i$\\
\hline
$\tilde{\beta}^i_{hv}$ &Probability that a recovered person infects a susceptible mosquito during a contact in patch $i$\\
\hline
$b^i_h(t)$  & Proportion of bites per human per unit time $t$ in patch $i$\\
\hline
$b^i_v(t)$  & Proportion of bites per mosquito per unit time $t$ in patch $i$\\
\hline
$\lambda^i_{vh}$  & Force of infection from vector to human  in patch $i$, \textit{i.e.,} measure of how likely a human get exposed in patch $i$\\
\hline
$\lambda^i_{hv}$  & Force of infection from human to vector  in patch $i$, \textit{i.e.,} measure of how likely a mosquito get exposed in patch $i$\\ 
\hline
\end{tabular}
\caption{Patches Parameters}
\label{tab:paramzone}
\end{table*}

Since our model is discrete, we model the transmission in patches at each discrete time step. 
For each patch $i$, we use almost the same parameters described in \cite{Chitnis2008} while adapting them in a multi-patches context (see Table \ref{tab:paramzone} for parameters details).

After identifying the required parameters, we define respectively the forces of infection vector-to-human ($\lambda^i_{vh}$) and human-to-vector($\lambda^i_{hv}$) in a patch as follows :
\begin{equation}
\lambda^i_{vh}(t+1) = b^{i}_{h}(t) \beta^{i}_{vh} \frac{I^{i}_{v}(t)}{N^{i}_{v}(t)} \hspace{0.9cm} 
\end{equation}

\begin{equation}
 \lambda_{hv}^i(t+1) = b^{i}_{v}(t)( \beta^{i}_{hv} \frac{I^{i}_{h}(t)}{N^{i}_{h}(t)} +  \tilde{\beta}^{i}_{hv} \frac{R^{i}_{h}(t)}{N^{i}_{h}(t)})
\end{equation}

%



\subsubsection{Individual mobility characteristics}
 \label{sec:mobility}   
\begin{figure}
     \centering
     \begin{subfigure}[b]{0.3\textwidth}
         \centering
         \includegraphics[width=\textwidth]{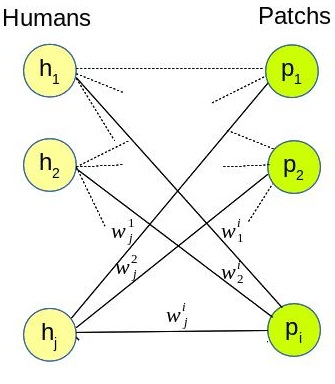}
         \caption{Individuals visiting different patches}
         \label{malaria_model}
     \end{subfigure}
     \hfill
     \begin{subfigure}[b]{0.3\textwidth}
         \centering
         \includegraphics[width=\textwidth]{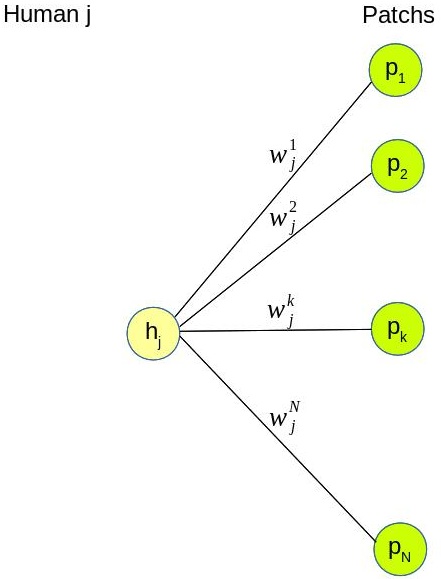}
         \caption{Mobility pattern of an individual}
         \label{fig:foiH}
     \end{subfigure}
     
      \caption{Human mobility and Patches}
        \label{fig:humanPatches}
\end{figure}

  We distinguish residential patches (cities or quarters) to \textit{ad-hoc} meeting patches. \textit{Ad-hoc} patches ($P_{M}$) are sparsely populated  and used as headquarter for social events while residential patches ($P_{R}$) are densely populated but not attractive for social meetings. Having this in mind, one may deduce that people move more often  from $P_{R}$ to $P_{M}$ than from $P_{R}$ to $P_{R}$. 
We assume that each individual is identified thanks to mobile applications and/or Telecommunication companies. Users data are anonymous in such a way that personal details are hidden while geographical positions of anonymous individuals are known at each time. 
 The time is discretized and we consider a sequence of consecutive time windows of equal duration. A time window $t$ is one night during which bites are much more frequent. At anytime, the patch of an individual $h_j$ and how long he stays there are  known. For instance, we observe on Fig \ref{malaria_model} and Fig. \ref{fig:foiH} that $h_j$ has stayed during $w_j^1$ time in $p_1$ and $w_j^2$ in $p_2$. 
 


The mobility within patches of an individual is used to compute the probability of $h_j$ to get exposed. 
The probability of getting exposed is function of the individual status and its FoI, which depends on its mobility as well as its immunity.

Basically, the FoI of an individual is the sum of the FoI of each visited patch $i$ weighted by its time presence in {\em i} : 
\begin{equation}
\lambda_{h,j}(t) = \sum\limits_{i=1}^N w^i_j(t)~\lambda^i_{vh}(t)
\end{equation}
Finally, the likelihood to get exposed of $h_j$ at time $t + 1 $ is : $pe_j(t+1) = pe_j(t) + \lambda_{h,j}(t) (1 - pe_j(t)).$
Once an individual is exposed, the rest of the incubation, development and recovery process is a matter of time. Basically, an exposed person has a likelihood of $p$ to get infected and {\bf to} recover after a certain period. Details of parameters for calculating the transition over the classes SEIR are described in Table \ref{tab:paramindiv}. 

\begin{table*}[t]
\centering
\rowcolors{2}{lightgray}{white}
\begin{tabular}{|c|p{10cm}|}
\hline
\textbf{Parameter} & \textbf{Description} \\
\hline
$h_j$ & Identification of the individual $j$ \\
\hline
$w_j^{k}$  & Visiting ratio of time that $h_j$ spends in patch $k$.
\\
\hline
$S_{h,j}(t)$ & Susceptible state variable of $ h_j $ at time step $ t $ (1 if susceptible 0 otherwise)\\
\hline
$E_{h,j}(t)$ & Exposed state variable of $ h_j $ at time step $ t $ (1 if exposed 0 otherwise)\\
\hline
$I_{h,j}(t)$ & Infected state variable of $ h_j $ at time step $ t $ (1 if infected 0 otherwise)\\
\hline
$R_{h,j}(t)$ & Recovered state variable of $ h_j $ at time step $ t $ (1 if recovered 0 otherwise)\\
\hline
$pe_j(t)$ & Probability that individual $h_j$ (being in class $S$) moves to class $E$
\\
\hline
$\nu_{h,j}$ &  
 incubation period (without symptoms). In the case of P. falciparum parasite, which predominates in Senegal, it varies from 9 to 10 days \cite{Chitnis2008,Cdc2015}.\\
\hline
$ \frac{1}{\gamma_{h, j}}$ & 
is the infectious period (Chitnis and Al. have set it at 9.5 months \cite{Chitnis2008})\\
\hline
\end{tabular}
\caption{Individual Parameters for a given time window}
\label{tab:paramindiv}
\end{table*}

After defining FoI of individuals, we identify at each time the new ones that will be exposed and let the model defines the persons that will occupy the remaining classes.

With the previous parameters, we may define the flowchart of the disease by gathering individual transitions.
To this end, we add the number of individuals belonging to each class for each patch in a regular basis.
Thus, individuals of a class are known as precisely located in a patch.   
The variation of the human population over the different classes are formalized as follows :

\begin{equation}
 S^i_h(t)= \sum\limits_{j=1}^M w^i_j(t)~ S_{h,j}(t) 
\end{equation}

\begin{equation} 
 E^i_h(t)= \sum\limits_{j=1}^M w^i_j(t)~ E_{h,j}(t) 
\end{equation}

\begin{equation}  
  I^i_h(t)= \sum\limits_{j=1}^M w^i_j(t)~ I_{h,j}(t) 
\end{equation}

\begin{equation}
   R^i_h(t)= \sum\limits_{j=1}^M w^i_j(t)~ R_{h,j}(t)
\end{equation}
   
The variation of mosquitoes population over the classes, are exactly the same as described in \cite{Chitnis2008} by the following equations:  
\begin{equation}
  	\frac{dS^i_v}{dt}= \psi^i_v N^i_v - \lambda^i_{hv}(t) S^i_v - f^i_v(N^i_v) S^i_v 
\end{equation}
  	
\begin{equation}
  \frac{dI^i_v}{dt}=\nu^i_v E^i_v - f^i_v(N^i_v) I^i_v
\end{equation}  	

\begin{equation}  	
   \frac{dE^i_v}{dt}= \lambda^i_{hv}(t) S^i_v - \nu^i_v E^i_v - f^i_v(N^i_v) E^i_v 
\end{equation}
where $f^i_v(N^i_v) = \mu_{1v}^{i} + \mu_{2v}^{i} N^i_v$  is the per capita density-dependent death rate for mosquitoes in the patch $i$ \cite{Chitnis2006}.

It is worth noting that after each time unit, we update data related to the patches and the individuals such as the FoI and the health status since they are used as input for the next time step.

\section{Malaria control using our model : A discussion}
 \label{sec:appli}
We discuss in this section some benefits of our model and how it can be used for the malaria control in southern part of Senegal. Basically, knowing mobility patterns of individuals {\bf provides}  additional informations for facing malaria. Among the various opportunities of the model, one may retain the following ones.
\begin{description}
\item \textbf{Risk customizing.} The model helps to compute the likelihood $p_E$ of a given individual to get exposed based on its mobility rate through the different patches over the time. Actually, a group of a patch population moving to another area can be made of a subgroup ($E_{moving}$) of exposed  individuals (respectively, susceptible and infectious), which differs from the exposed others ($E_{staying}$) of the same patch. That is, the mean of the probability $p_E$ of moving individuals is not the same than the $p_E$ of the overall population of the same patch. In a context like Kedougou, such an observation states that truckers, traders, and others with a high mobility rate should be observed with a distinctive approach since their risk to get infected, with  the disease dispersal knock-on effect, is more significant. 
\\

\item \textbf{Reducing antimalarial costs.}
Since the model is devised for each individual, 
it allows to target specific persons at higher risk than the overall population. Bearing this in mind and the fact that each antimalarial action costs, then using the model can contribute to reduce the necessary means  for the surveillance, and eventually, the elimination of the disease. Moreover, with the tight budgets in developing countries, combined with the rapid growth of the demography as well as the explosion of other infectious diseases that create new priorities for governments, the proposed approach seems to arrive at the right time for facing definitively against malaria.  
\\

\item \textbf{Mobility impact.}
As shown in the validation section,  mobility may have either a great or low impact on a visited patch. Therefore, the time-based follow-up of a patch's FoI using current individuals mobility gives enough room for producing more clinical results in terms of predictions. In opposite of current strategies applied in Kedougou, which do not include human mobility details and discover the overall disease trends weeks or months later, our model plots instantly detailed information of the malaria dispersal. Hence, policy makers may see the right actions to do in each patch even though the disease not has not yet happened within.
\\

 \item \textbf{What-if Analysis boost.}
 Last but not the least, our solution can be used to calibrate the overall actions against malaria. In fact, we can model the disease spread while asking or supposing a specific pattern. For instance, we can suppose (or eventually suggest) that people have to stay home during their incubation period  in order to reduce the global evolution of the disease. Likewise, we can suggest them following a specific pattern based on the FoI of different patches.  
 In other words, we offer to policy maker a tool that can be used as dashboard to evaluate different scenario and their effects. 

\end{description}

\vspace{-0.3cm}
\section{Implementation and Validation}
\label{sec:exp}

\subsection{Experimental Setup}
We implemented our approach using the version 2.7.15 of Python through Spyder IDE 3.2.6 on Linux. We rely on Jupyter for visualization  and share the source code of our implementation on GitHub~\footnote{https://github.com/Klynda/Prevention-du-paludisme}.
To calculate the model, we use many parameters for zone's characterization and individuals too. All these parameters are detailed in section \ref{characteristics}. 
We recall that most of these parameters values have been reported  in the literature. 

We consider two patches : the residence zone $P_R$ and the meeting one $P_M$, with their respective human population  $|P_R|$ and $|P_M|$, their respective vector (mosquito) population $NV_R$ and $NV_M$ and the proportion $p$ of human traveling from $P_R$ to $P_M$. We observe $I_R$, the number of infected persons in $P_R$.
Table~\ref{tab:param} summarizes the parameters used in the experiments.

\begin{table*}[htbp]
\centering
\caption{Experimental parameters}
\begin{tabular}{| l | l | l |}
\hline
name 	 	& description &  value\\
\hline
$P_R$ 	 	& the residential patch & \\
$|P_R|$	 	            & the number of individuals in $P_R$ & 3000\\
$NV_R$ &  the number of vectors in $P_R$ & Varying \\
$P_M$ 	 	& the meeting patch &\\
$|P_M|$	 	            & the number of individuals in $P_M$ & 1000\\
$NV_M$ &  the number of vectors in $P_M$ & Varying \\
$p$         	    & the ratio of moving individuals ($p>0$) & $[0.1, 0.4]$\\
$I_R$         	    & the number of infected individuals in $P_R$ & [0, 3000]\\
$\psi$         	    & vector birth rate  & 0.03\\
$\mu$         	    & vector death rate  & [0.03, 2E-8] \\
\hline
\end{tabular}
\label{tab:param}
\end{table*}
%

\subsection{Experimental Objectives and Method}
The overall goal of the validation is to investigate the benefit of individual mobility for malaria control. 
We intend to show that taking into account individual mobility allows for a more accurate modeling of the disease evolution over time and space (\ie patches). 
We aim to simulate disease evolution which cannot be captured by existing models that are unaware of individual mobility. 
Precisely, we evaluate the gain of our approach in three different aspects: 


\begin{enumerate}
 
\item The impact of individual mobility on the estimation of infected individuals.

\item The relevance of the proposed model to  approximately match  recently reported real cases.

\item The vector control opportunities based on individuals movements and patches characteristics.

\end{enumerate}

\subsubsection{Size of the vector population over seasons}


In this section, we only consider the $P_R$ patch, thus we omit the $R$ indice in the notations.
To be as close as possible to what happens in the real world, we vary the population of (mosquito) vectors according to the two main seasons occurring in the Kedougou region: 

\begin{itemize}
\item the rainy season (approximately from start of June to the end of November) 
\item the dry season (the remaining 6 months).
\end{itemize}

The number of mosquitos grows \wrt a daily birth rate ($\psi$, see Section~\ref{tab:param}). This birth rate still grows and reaches a stationary value when the rainy season settles definitively. Indeed, the mosquito birth rate is correlated with the amount of wet place (reproduction areas). All potential wet places are full of water when the rainy season is in full swing. Thus, we assume that during the rainy season the total size of these wet areas remains almost constant therefore the vector population $NV_{max}$ remains almost constant.

Then, at the end of the rainy season, the vector population is gradually
decreasing up to the dry season ceiling (i.e. a rather small number that makes the epidemic stop itself). The death rate ($\mu$  see Table~\ref{tab:param}) is based on the average life duration of a vector (30 days).

\begin{figure}[htbp]
\centering
\begin{subfigure}[b]{\linewidth}
         \centering
         \includegraphics[width=\linewidth]{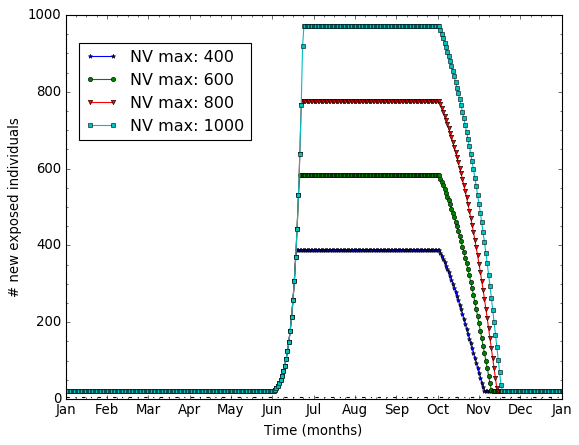}
         \caption{NV max varying from 400 to 1000}
     \end{subfigure}
 \begin{subfigure}[b]{\linewidth}
         \centering
         \includegraphics[width=\linewidth]{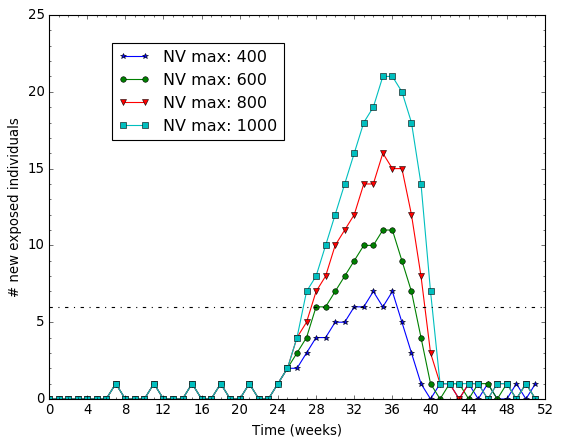}
         \caption{\# new exposed indiv. vs. NV max}
     \end{subfigure}    
 
\caption{Impact of the maximum number of vectors ($NV_{max}$) on the disease}
 \label{fig:Nv}    
\end{figure}

Figure~\ref{fig:Nv}(a) plots the number of vectors over time during one year for the residential patch, with different characteristics in terms of wet areas. For instance, in the first case ($NV=400$) the vector population is 20 times higher than during the dry season, whereas in the last one, ($NV=1000$) it is 50 times higher.
This allows for simulating areas with different characteristics in terms of wet areas and therefore in terms of vectors population growth. Note that these areas are located in the same region thus have {\bf similar} rainy seasons (from June to November). 

%

To figure out the effect of the rainy season on the disease, we compute the number of newly infected individuals in $P_R$ patch when everyone remains sedentary (\ie no mobility).
Figure~\ref{fig:Nv}(b) reports the results for the 4 patches from  Figure~\ref{fig:Nv}(a). In the first patch ($Nv=400$) there is almost no disease while in the 3 other ones, the peak disease grows 
for patch having more vectors. 
We use these preliminary simulations, to set the maximal number of vectors (dented $Nv_{max}$) that a residential and meeting patch have in the subsequent experiments. 
This allows us to define a low endemicity residential patch ($Nv_{max}=400$) and a higher endemicity meeting patch ($Nv_{max}=1000$)

\subsection{Impact of individual mobility} 

The goal of this section is to quantify the impact of individual mobility in modeling malaria. 
Based on  the Kedougou case ({\em cf.} Section~\ref{sec:intro}), we consider a village of farmers that sell their products at a remote market. 
Basically, there are two patches: a residential place $P_R$ 
and a market place $P_M$.
There are 3000 people living in $P_R$. Among them, a group of people (which size ratio is $p$ relative to $P_R$ population) moves everyday from $P_R$ to $P_M$ and come back home.

\subsubsection{Varying the mobility rate}
\begin{figure}[htbp]
    \centering
        \includegraphics[width=\linewidth]{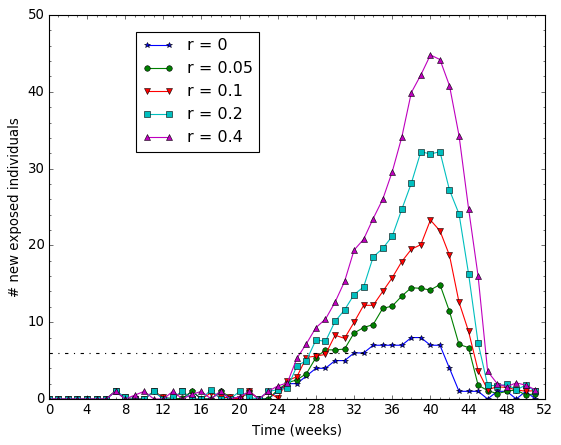}
        \caption{Varying the mobility rate $r$ from 0 to 40\%}
        \label{fig:moving}
\end{figure}

The goal of this experiment is to assess the impact of the mobility on the disease evolution. First, we define the \textit{mobility rate} $r$ as the ratio of  people moving from $P_R$ to $P_M$.
Then we investigate how the number of exposed individuals, $E(r)$,  evolves over time for various mobility rates. 
To this end we vary the mobility rate, $r$, from 0\% to 40\%. On Figure~\ref{fig:moving}, we report the number of newly infected individuals per week. 

The dashed black line indicates the threshold limit of exposed people. Above this threshold, the disease is qualified as an epidemic situation. The threshold value is set to 6 new cases per week according to real observations reported in Kedougou during years 2016 and 2017 \cite{sn:gouv:sante}.  

We observe on Figure~\ref{fig:moving} that the higher the moving rate, the higher the epidemic and longer is the epidemic duration too. For example, when nobody is moving (r=0), the epidemic lasts 8 weeks with 7 new cases per week; 
whereas for a moving rate of 10\%, it lasts 16 weeks and reaches a peak of 25 cases per week.

\subsubsection{Early/Late rainy season in the meeting patch $P_M$}
\label{early_late}
In case the mobility patch and the residential patch have slightly different raining seasons, this could have an impact on the epidemic duration.\\
In this section, we study the impact of having a rainy season in $P_M$ that does not start exactly at the same time as in $P_R$.
The rainy season in $P_R$ lasts from week 22 to week 38 (\ie 16 weeks from June to September).

\begin{figure}[htbp]
\centering
\begin{subfigure}[b]{\linewidth}
         \centering
         \includegraphics[width=\linewidth]{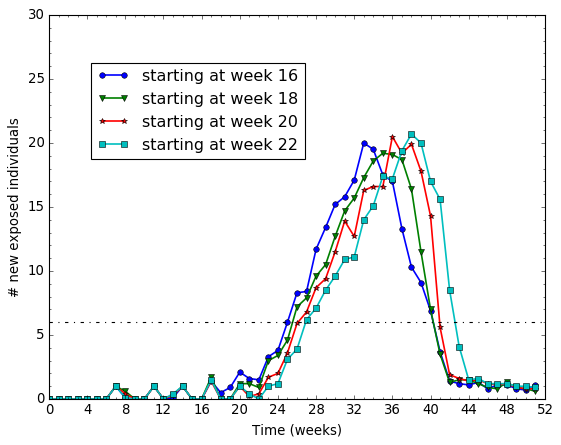}
         \caption{Starting in $P_M$ \textbf{earlier} than in $P_R$}
     \end{subfigure}
 \begin{subfigure}[b]{\linewidth}
         \centering
         \includegraphics[width=\linewidth]{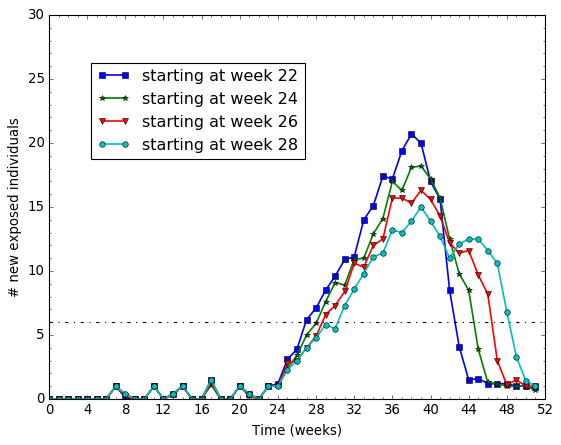}
         \caption{Starting in $P_M$ \textbf{later} than in $P_R$}
     \end{subfigure}    
 
\caption{Varying the starting date of the rainy season in $P_M$}
 \label{fig:delay}    
\end{figure}

The $P_M$ rainy season lasts as long as the $P_R$ one but it starts before or after week 22.
The mobility pattern is set to 20 \% mobile individuals that go to $P_M$ every day for  half of their time.
We report on Figure~\ref{fig:delay} the number of exposed individuals for several starting dates of the $P_M$ rainy season.

The results show that the epidemic lasts longer in figure \ref{fig:delay}(b)  than in figure \ref{fig:delay}(a). That is, it lasts longer when the rainy season starts in $P_M$ later than in $P_R$. The extra time duration (between 2 and 6 weeks as reported in Figure \ref{fig:delay}(b)) of the epidemic corresponds to the difference at the starting time of rainy season for the two patches.

\subsubsection{Varying the mobility starting date}

In this experiment, we vary the starting date (before this date, nobody moves) from t=0 (beginning of the year) to t=40 (late October). We want to investigate the impact of seasonal migration on the development of the disease. Intuitively, we expect the greater impact if the migration occurs during the rain season, when the vector population is at its maximum. Such migrations are usual in the Kedougou region, where there are few fair places and people from small cities or villages have to move to sell or buy goods.

\begin{figure}[htbp]
\centering
\begin{subfigure}[b]{\linewidth}
         \centering
         \includegraphics[width=\linewidth]{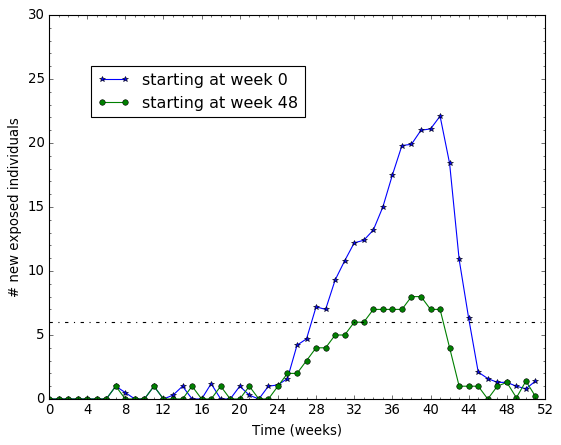}
         \caption{Mobility starting before/after the rain season}
     \end{subfigure}
     \vfill
 \begin{subfigure}[b]{\linewidth}
         \centering
         \includegraphics[width=\linewidth]{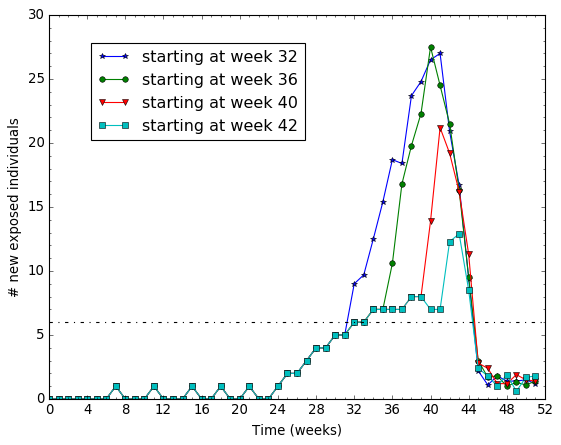}
         \caption{Mobility starting during the rain season}
     \end{subfigure}    
 
\caption{Impact of the mobility starting date on the disease}
 \label{fig:debMob}    
\end{figure}

\begin{figure}[htbp]
    \centering
        \includegraphics[width=\linewidth]{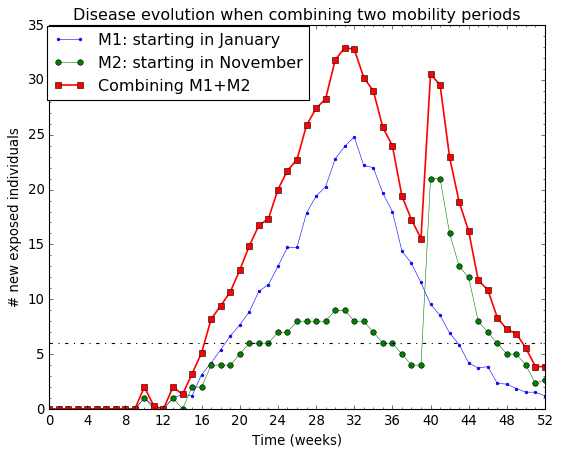}
        \caption{Aggregating on 2 zones}
        \label{fig:debMobSomme}
\end{figure}

With 20\% moving people and other parameters set similar to those ones used in previous experiment (section \ref{early_late}), we plot on Figure~\ref{fig:debMob} the impact of the starting mobility on the disease evolution. \\
The results (for "Week 40" curve) suggest that the disease development is quite slow before the migration start. 
When the migration starts at the beginning of the rain season (see "Week 0" curve), the disease grows slowly because most of the vectors are yet in susceptible state implying a low vector-to-human FOI.
On the other hand, when a migration starts at the middle of the rain season (see "Week 32" and "Week 36" curves), the disease grows very fast because most of the vectors are  already infected, thus, causing a high FOI.\\
Figure~\ref{fig:debMobSomme} aggregates 2 different cases of migration patterns occurring on 2 sub-areas: a migration starting at week 0 and another one starting at week 40. We report (in red curve) the total number of newly exposed people on the area.\\
The results suggest that the disease lasts 35 weeks, which is longer than any of the two sub-areas. More interestingly, the disease lasts 8 weeks longer than the longest epidemic plotted on Figure \ref{fig:moving}. Notice that  the two sub-areas do have the same rain season because they are located in the same region.
We can conclude that successive migrations from various specific areas (close villages) tend to generate rather long epidemic at a higher scale (region level).
The results are consistent with the real Kedougou observations : they provide a possible explanation of what happens at Kedougou.

\subsection{Relevance of the model to match the Kedougou real case}
The objective is to evaluate the relevance of our model to match real observations recently reported surveillance bulletins in~\cite{pnlp:kedougou:2017}.
An (malaria) observation is reported as a series of newly infected people values, one value per week. 
Given a observation occurring in an area of $P$ inhabitants for a period of $n$ weeks, we define a normalized report $R = \{R_1, \cdots, R_n\}$ such that $R_i$ is the number of newly infected people for week $i$ divided by the population $P$.
%
Let $M$ be a model for the observation reported by $R$. 
Running $M$ generates $\{M_1, \cdots, M_n\}$ such that $M_i$ is the expected ratio of newly infected people on week $i$.
%
  
\begin{figure}[htbp]
    \centering
        \includegraphics[width=\linewidth]{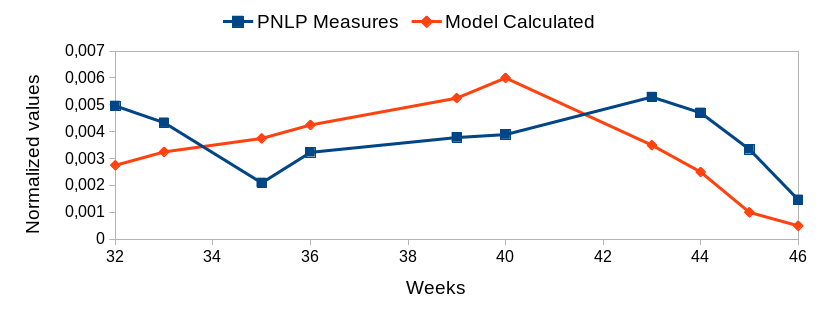}
        \caption{Normalized values from observation (reports) and from the Model (calculated)}
        \label{fig:normailzedVal}
\end{figure}

 We plot the obtained values in Figure~\ref{fig:normailzedVal} and the relative accuracy is $E_{M,R} = 0.016$, what gives a mean absolute error $MAE=0.001$. These results show that the values measured (reports) and those calculated with the model differ by approximately 1 case per 1000.  Therefore we can say that our model produces values that are close to what is reported from real observations. 
 

Therefore, using our model can help for more efficient malaria control actions. For example, using the model one can decide which location is priority when it comes to do some preventives actions. 


\subsection{Vector control efficiency} 
In this experience, we measure the benefit of our approach on controlling the malaria vector.
There are two types of malaria control actions: 1) a \emph{preventive action}, which consists of convincing someone to use repellent and mosquito nets to avoid mosquito bites, and 2) an \emph{eradication action} (\ie mosquito removal) that consists in suppressing most of the vectors in an area using chemical products. Notice that this second action type may have dramatic ecological consequences. Therefore, the first action type, preventive, would be a better choice. However, this action has a cost that must be optimized. 
 

We aim to show that preventive action helps reducing the FoI vector-to-human and then reduce the epidemic intensity and duration.

\begin{figure}[hptb]
\centering
\includegraphics[width=0.7\linewidth]{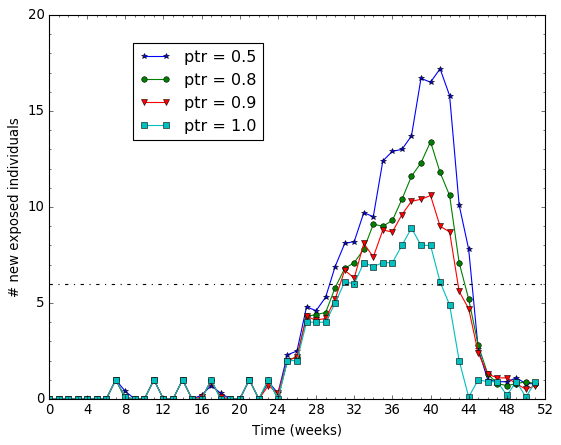}
\caption{Malaria evolution according to the mosquito control strategy}
\label{fig:exp55_results}
\end{figure}

We already show in previous sections, that the moving part of the population is the major factor that impact the epidemic duration and intensity. We now protect those people who move from residential patch to meeting patch. Actually, this protection could be done through repellent and mosquito nets.   
In this respect, we consider the experiment configuration where individuals mobility rate is 10\% (see figure \ref{fig:moving} when $r=0.1$ ) and we use use different values of protection rate ($ptr$) for those people who move regularly. The protection rate $ptr$ depicts the ratio of people among the moving ones who are protected. The results of these experiments are shown on figure ~\ref{fig:exp55_results}. 

%

We show that for $ptr=1$ when we protect 100\% of the moving population, the epidemic intensity and duration is as low as if no people is moving (see the $r=0$ case in figure \ref{fig:moving}). 
Therefore, preventive actions targeting moving people can be rather efficient.
In a residential patch where few people are moving, such preventive action would be cost-optimized assuming that the cost per individual of a preventive action is low. 
When $p$ is varying from $1$ to $0.8$, the total number of exposed individuals is growing respectively from $118$ to $169$. This means that the preventive action must be rather complete to be efficient.

\vspace{-0.4cm}

\bibliographystyle{apalike}
{\small
\bibliography{biblio}}

\end{document}